\documentclass{article}

\usepackage{PRIMEarxiv}

\usepackage[utf8]{inputenc} 
\usepackage[T1]{fontenc}    
\usepackage{hyperref}       
\usepackage{url}            
\usepackage{booktabs}       
\usepackage{amsfonts}       
\usepackage{nicefrac}       
\usepackage{microtype}      
\usepackage{lipsum}
\usepackage{fancyhdr}       
\usepackage{graphicx}       
\graphicspath{{media/}}     

\title{Urban precipitation downscaling using deep learning: a smart city application over Austin, Texas, USA
}

\author{
  Manmeet Singh \\
  Jackson School of Geosciences \\
  University of Texas at Austin \\
  Austin\\
  \texttt{manmeet.singh@utexas.edu} \\
   \And
  Nachiketa Acharya \\
  CIRES, University of Colorado Boulder \\
  NOAA/Physical Sciences Laboratory \\
  Boulder\\
  \texttt{nachiketa.acharya@noaa.gov} \\
   \And
  Sajad Jamshidi \\
  Department of Agronomy \\
  Purdue University \\
  West Lafayette\\
  \texttt{sjamshi@purdue.edu} \\
  \And
  Junfeng Jiao \\
  Department of Civil, Architectural, and Environmental Engineering, Cockrell School of Engineering \\
  University of Texas at Austin \\
  Austin\\
  \texttt{jjiao@austin.utexas.edu} \\
\And
  Zong-Liang Yang \\
  Jackson School of Geosciences \\
  University of Texas at Austin \\
  Austin\\
  \texttt{liang@jsg.utexas.edu} \\
  \And
  Marc Coudert, Zach Baumer \\
  Office of Sustainability \\
  City of Austin \\
  Austin\\
  \texttt{\{marc.Coudert, zach.Baumer\}@austintexas.gov} \\
    \And
  Dev Niyogi \\
  Jackson School of Geosciences \\
  University of Texas at Austin \\
  Austin\\
  \texttt{dev.niyogi@jsg.utexas.edu} \\
}

\begin{document}
\maketitle

\begin{abstract}
Urban downscaling is a link to transfer the knowledge from coarser climate information to city scale assessments. These high-resolution assessments need multiyear climatology of past data and future projections, which are complex and computationally expensive to generate using traditional numerical weather prediction models. The city of Austin, Texas, USA has seen tremendous growth in the past decade. Systematic planning for the future requires the availability of fine resolution city-scale datasets. In this study, we demonstrate a novel approach generating a general purpose operator using deep learning to perform urban downscaling. The algorithm employs an iterative super-resolution convolutional neural network (Iterative SRCNN) over the city of Austin, Texas, USA. We show the development of a high-resolution gridded precipitation product (300 m) from a coarse (10 km) satellite-based product (JAXA GsMAP). High resolution gridded datasets of precipitation offer insights into the spatial distribution of heavy to low precipitation events in the past. The algorithm shows improvement in the mean peak-signal-to-noise-ratio and mutual information to generate high resolution gridded product of size 300 m X 300 m relative to the cubic interpolation baseline. Our results have implications for developing high-resolution gridded-precipitation urban datasets and the future planning of smart cities for other cities and other climatic variables.
\end{abstract}


\section{Introduction}
The past decade has witnessed a remarkable growth in urbanization because of human migration and economic developments. Decisions about city planning, operations, hazard response and recovery, and climate resiliency require high spatiotemporal resolution environmental information. Events such as flash floods, for instance, are caused by heavy precipitation and data are needed to develop the climatological understanding that can help mitigate smart approaches that can minimize losses to property and humans. Thus, a number of city based operations need high resolution climate information. Currently the City of Austin is developing a climate projection for the city that can be used for different sustainability operations. For this purpose, the data available is typically from reanalysis or satellite gridded fields and that needs to be downscaled. The location of Austin is shown in Figure. 1. As the capital of the state of Texas, United States state and the headquarters and largest city of Travis County, Austin has a population of more than 500,000 people. For a city of its size, Austin is a big one. It's the 11th-largest in the United States of America, the 4th-largest in the state of Texas, and the second-largest state capitol after Phoenix, Arizona. Since 2010, it has been one of the fastest-growing big American cities. Austin's population was estimated at 961,855 people in the most recent census in 2020. Lady Bird Lake and Lake Travis on the Colorado River, Barton Springs, McKinney Falls, and Lake Walter E. Long that fall within Austin in Central Texas are just a few of the many lakes, rivers, and waterways that make up the Texas Hill Country. High-resolution downscaled urban data products are essential for the city of Austin to plan for the future climate. The study domain is also motivated by our knowledge of the urban morphology, local datasets, and stakeholder needs. For example, the City of Austin is keen on identifying hotspots of air pollution, regions at risk of flooding \cite{zhou2019high}, and locales of excess heat \cite{wang2018hyper} are of interest to the study domain for which high-resolution data implications over urban regions. The input data are based on publicly available spatial rainfall product that have been recently evaluated for their accuracy over the land areas \cite{you2020comparison}. The final downscaled output resolution at 300 m is also guided by the end user needs posed for developing city-scale climate resiliency studies for sub-km scale datasets.

\begin{figure}[t]%
\begin{center}
\includegraphics[width=0.55\textwidth]{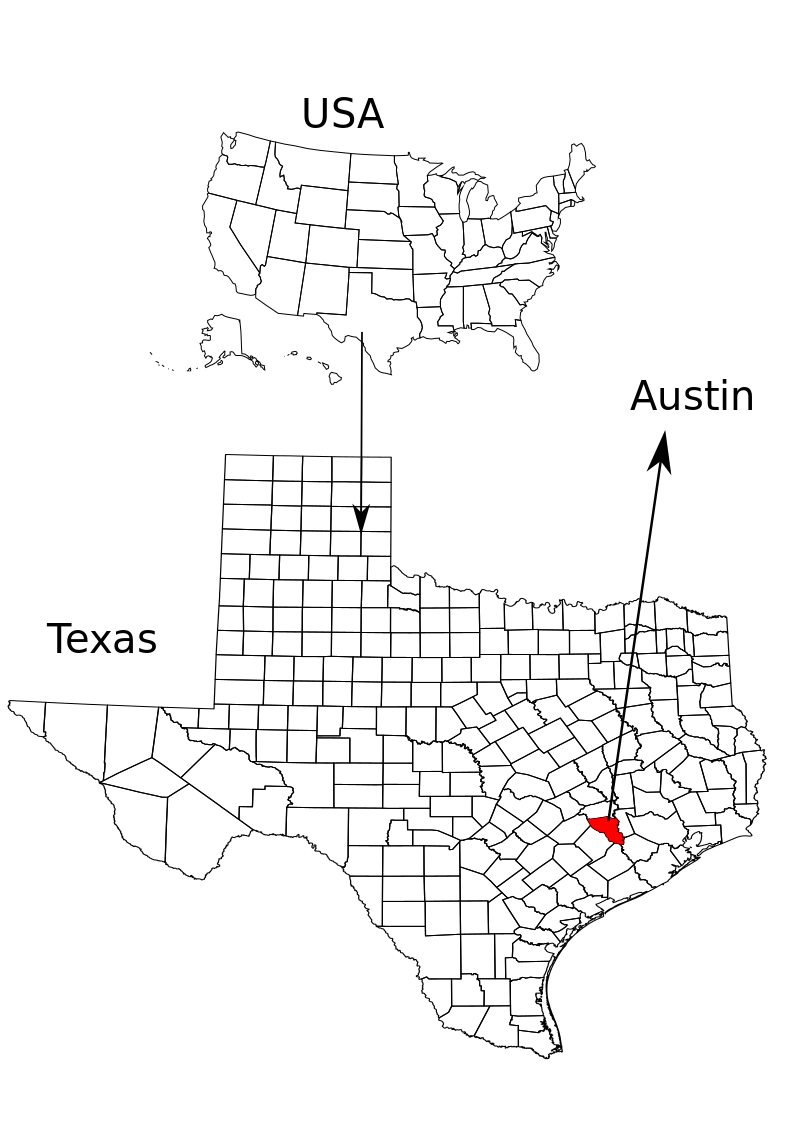}
{\caption{Location of Austin, Texas in the USA. Urban downscaling is performed over a 3 $^{\circ}$ X 3 $^{\circ}$ box (29-32N,96-99W) centered over Austin.}
\label{fig1}}
\end{center}
\end{figure}

Development of the urban precipitation high resolution data/ climatology is important for a variety of planning as well as water resources and disaster response activities. In addition, such information is important for the development of infrastructure as part of the smart city framework (\cite{anthopoulos2017smart}). Cities are seeking to develop climate resiliency and sustainability strategies for which urban scale (i.e., spatial scale order of 1km grid spacing) and climatic datasets are needed. There are popular climate data available from international and coordinated assessments that have resulted in the Reanalysis and climate model outputs (e.g., IPCC/CMIP6 or ERA5); however, their grid spacing is relatively coarse (order of 10s to 100s of km grid).  Projecting the future state of the atmosphere has been made possible using numerical models (\cite{yang2016structure}). Despite significant improvements to the numerical models in the last decade, the limitation in the computational power and numerical stability (\cite{navon2009data}) mean the global climate model and reanalysis outputs are generated at coarse spatial resolution. While adequate for modeling mesoscale processes and weather forecasting (\cite{sha2020deep}), this resolution is not optimal for capturing spatial variability of environmental and climate variables in a heterogeneous environment and complex terrains (\cite{schumacher2020formation}).

As a result, cities are generally represented by a single or similar small number of grids from the climate reanalysis fields. The climatology that emerges from such large scale fields is of limited use for city-scale operations requiring information at a much higher spatio temporal resolution. City departments need such information to understand local vulnerabilities, assess infrastructure planning needs, and develop resiliency plans considering equity and adaptive options available at their disposal. Additional examples of the high-resolution analysis include working with problems such as water and food security, dealing with infectious disease and heat, air quality long-term exposure assessments, and developing energy and other demand studies.
\begin{table}[t]
\tabcolsep=3pt
\caption{Comparison of machine learning / deep learning studies focused on urban downscaling}\label{tab1}%
\begin{tabular}{@{}lllllll@{}}
\toprule
\textbf{A} & \textbf{B} & \textbf{C} & \textbf{D} &
\textbf{E} & \textbf{Downscaling input \& target}  & \textbf{Study} \\
\midrule
\textbf{LST} & Guangzhou & RF & x  & x  & 90 m to 10 m & \cite{gu2015recent}  \\
\textbf{LST} & Seoul & DL & x & x  & 1000 m to 30 m & \cite{xu2020downscaling}  \\
\textbf{Tb} & Guangzhou & RF & x & x  & 30 m to 10 m & \cite{choe2020improving}  \\
\textbf{LST} & Zhangye & RS & x & x  & 270 m to 90 m & \cite{xu2021hybrid}  \\
\textbf{LST} & Jordan river valley & RF & x & x  & 1000m to 250 m & \cite{pan2018applicability}  \\
\textbf{LST} & Los Angeles & SVM & x & x  & 5km to 1km & \cite{hutengs2016downscaling}  \\
\textbf{LST} & Beijing & SVM, RF, ANN & x & x  & 990m to 90m & \cite{weng2014modeling}  \\
\textbf{NO2} & Los Angeles & DL & x & x  & 0.125$^{\circ}$ to ~ 5km & \cite{li2019evaluation}  \\
\textbf{PM2.5} & USA & RF+KR & x & x  & 0.1$^{\circ}$ to 0.01$^{\circ}$ & \cite{yu2021deep}  \\
\textbf{Tmax/Tmin} & USA & RF & x & x  & 0.25$^{\circ}$ to 4 km & \cite{liu2018improve}  \\
\textbf{T2m} & Tokyo & DL & \checkmark & x  & 10m to 5m & \cite{sha2020deep}  \\
\textbf{Pr}& Austin, Texas, USA & DL & \checkmark & \checkmark  & ~10 km to 300 m & SRCNN (This study)  \\

\hline

\end{tabular}

Abbreviations: \textbf{A}: Variable (LST:Land surface temperature; Tb: Brightness temperature; PM2.5: Particulate Matter smaller than 2.5 micron; Pr: precipitation; NO2: Nitrogen; Tmax/Tmin: maximum/minimum 2-m temperature; T2m: Two meter air temperature), \textbf{B}: City/Location(GZ:Guangzhou; ZH:Zhangye; SL: Seoul; JRV:Jordan river valley; LA: Los Angeles; BJ: Beijing), \textbf{C}: (RF: Random forests; RS: Remote sensing indices; DL: Deep Learning; SVM: Support Vector Machine; ANN: Artificial Neural Networks; KR: Kriging), \textbf{D}: Convolutional Neural Networks, \textbf{E}: Iterative downscaling

\end{table}

\begin{figure}[t]%
\includegraphics[width=1.0\textwidth]{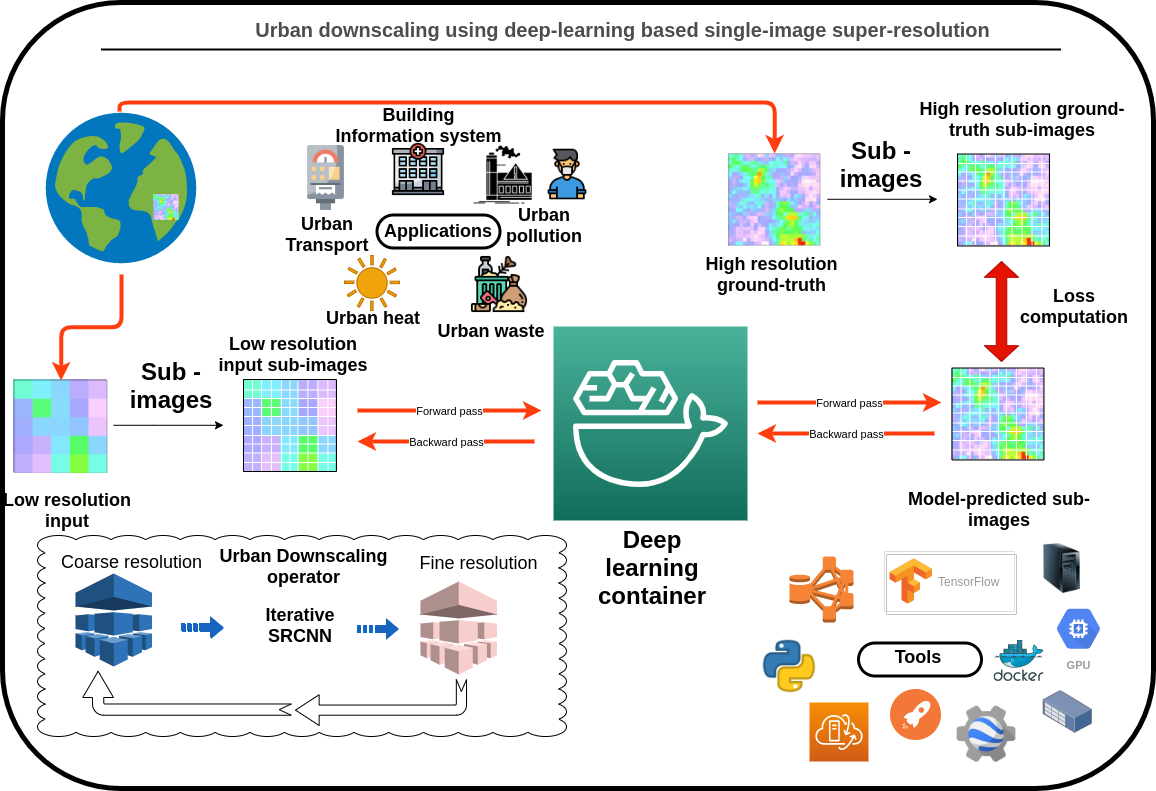}
{\caption{Schematic showing the process used to downscale the rainfall dataset over Austin, Texas, USA using SRCNN in this study. }
\label{fig1}}
\end{figure}

A data-driven decision narrative is often needed for cities to develop smart solutions as part of their operational efficiency, improved livability and short- and long-term resiliency outlooks. There is an increasing demand for high spatiotemporal resolution data over the urban regions for smart growth planning, emergency response, and management in response to the current changing climate (see \cite{holden2011empirical}, \cite{liu2020high}). The rainfall and clouds over urban areas vary due to anthropogenic activities and changes in the land use/land cover characteristics. In the study by (\cite{freitag2018urban}), urban imprints were found in the precipitation and cloud processes in addition to changes caused to the upstream flow of water. Extreme rainfall over urban areas and particularly over urban-rural boundaries has shown increasing trends, and the signature can be found across the world (\cite{freitag2018urban}, \cite{niyogi2017urbanization}). The characteristics of storms producing flash floods over urban areas were studied by (\cite{kishtawal2010urbanization}). A resilient and sustainable response to the current and future climate scenarios relies on an accurate understanding of how climatic characteristics are modified by different sub-sections of a city. Accordingly, researchers have generated surface flux data at the sub-city scale resolutions (primarily for surface temperature and air quality) using different downscaling approaches (e.g., \cite{agathangelidis2019improving}, \cite{hofierka2020physically}). 

A downscaled urban precipitation product at high spatiotemporal scales is necessary to capture the different active processes and becomes a high-value product. To circumvent the coarse-scale issue for impact studies, downscaling approaches have been employed to generate high spatial resolution data. Downscaling is the process of improving the resolution of coarse grid and sampling frequency datasets to higher resolution outputs. The operators used for such a transformation range from computationally expensive downscaling models (e.g.,\cite{leung2005downscaling}) to the simpler two-dimensional linear interpolation (\cite{shepard1968two}). Several statistical techniques have been applied in the literature related to cubic interpolation, kriging methods, random forests, support vector machines, artificial neural networks, and deep learning-based approaches (\cite{sun2020downscaling}, \cite{sekulic2021high}, \cite{sha2020deep}, \cite{sachindra2018statistical}, \cite{wang2021deep}). Downscaling approaches are postprocessing techniques that can be categorized under two overarching themes of statistical and dynamical approaches. An example would be developing a relationship using a large-scale climate product and then assuming that relation holds at a local scale and generating high-resolution fields. Recently more sophisticated statistical approaches are available: for example, change detection method (\cite{hu2019satellite}), Support Vector Machine-Probabilistic Global Search (\cite{njoku2002observations}), and artificial intelligence (AI). These statistical approaches have gained popularity in recent years due to their ability to upscale and downscale meteorological parameters (e.g., K-Means, Neural network) and due to the relatively quick execution and computational needs compared to the dynamical downscaling methodology. Convolutional neural networks (CNNs) are AI-based algorithms (i.e., deep learning) that consist of a series of convolutional layers that: (i) slide along inputs (as multidimensional arrays), (ii) assigns learnable weights, and biases to each neuron, and (iii) generates the featured output map (\cite{ghosh2010svm}, \cite{aloysius2017review}). Given CNNs ability in learning the patterns from gridded datasets, they have been used in several downscaling approaches (e.g., \cite{gu2015recent}, \cite{xu2020downscaling}).

With the surge in recent deep learning/ machine learning techniques, there is growing evidence that the traditional statistical methods used to increase the spatial resolution or downscaling can be enhanced by deep learning-based techniques. More specifically, the work by (\cite{dong2015image}), showed that a simple, lightweight image-to-image deep convolutional neural networks (SRCNN) outperform a widely used technique for spatial downscaling: the two-dimensional cubic interpolation substantially. They also note that the simple deep learning-based solution is comparable to the sparse coding technique. Another study by (\cite{kumar2021deep}) found that SRCNN improved spatial downscaling for a large region such as South Asia. Furthermore, incorporating additional variables such as topography and stacking SRCNN could additionally enhance the downscaled output (\cite{vandal2017deepsd}). Table 1 lists studies that have employed CNNs-based algorithms (i.e., random forest, Remote sensing indices, Deep Learning, support vector machine; artificial neural networks) to generate higher spatial resolution data over urban regions. Various studies (\cite{gu2015recent}, \cite{xu2020downscaling}, \cite{choe2020improving}, \cite{xu2021hybrid}, \cite{hutengs2016downscaling}, \cite{weng2014modeling}, \cite{li2019evaluation}, \cite{yu2021deep}, \cite{liu2018improve}, \cite{sha2020deep}) have explored downscaling variables important for urban areas. Examples include downscaling of land surface temperature, air temperature, air pollutants such as fine particulate matter (PM2.5), and nitrogen dioxide (NO2). These studies have mostly focused on cities in the United States of America, China, Japan, and South Korea. The past studies (Table 1) have focused on machine learning methods such as Random Forests, kriging, support vector machines, and artificial neural networks. However, the application of deep learning is a recent phenomenon. While most of these studies attempt to improve the spatial resolutions of the urban datasets, they do not employ CNN, which have shown superior performance on image-based tasks. Moreover, the past studies have primarily focused on temperature and air pollution-related variables, and high-resolution precipitation maps over urban regions, which are critical for urban hydrology applications, have not been assessed by any study. Moreover, as shown in Table 1, these studies have mostly attempted at the downscaling factors of up to 10x, except the work done by (\cite{xu2020downscaling}), who attempt to downscale up to 30x of the low-resolution inputs.

In this study, we focus on the problem of precipitation downscaling over urban areas using the deep learning/SRCNN approach using Austin, Texas, USA as the urban domain. While applications of CNN-based methods have resulted in satisfactory outcomes, the networks of these algorithms are rather complex. Super-Resolution Convolutional Neural Network (SRCNN)  (\cite{dong2015image}) is a simple light-weight network structure and has a high restoration quality. Given the potential of the SRCNN method, we postulate that high-quality rainfall data with fine spatial resolution could be generated using iterative SRCNN over the urban region with higher accuracy and speed than regular CNN methods. We test this approach because, downscaling rainfall data over the urban areas is particularly challenging as the rainfall characteristics are modified by the city’s microclimate (\cite{freitag2018urban}). The heterogeneous environment of urban regions with varying physical and thermodynamic properties and anthropogenic activities alters the surface flux and impact the atmospheric boundary layer, which ultimately translates into a shift in rainfall regime over the urban landscape (\cite{onishi2019super}).  We test this hypothesis over the urban region of Austin, Texas by ingesting data from the Japan Aerospace Exploration Agency (JAXA) satellite product as a precursor to the deep learning model. Our method is expected to create an avenue for generating high spatiotemporal meteorological datasets with super-resolution that have significant implications in the current and future development plans of smart cities.
\begin{figure}[t]%
\begin{center}
\includegraphics[width=0.55\textwidth]{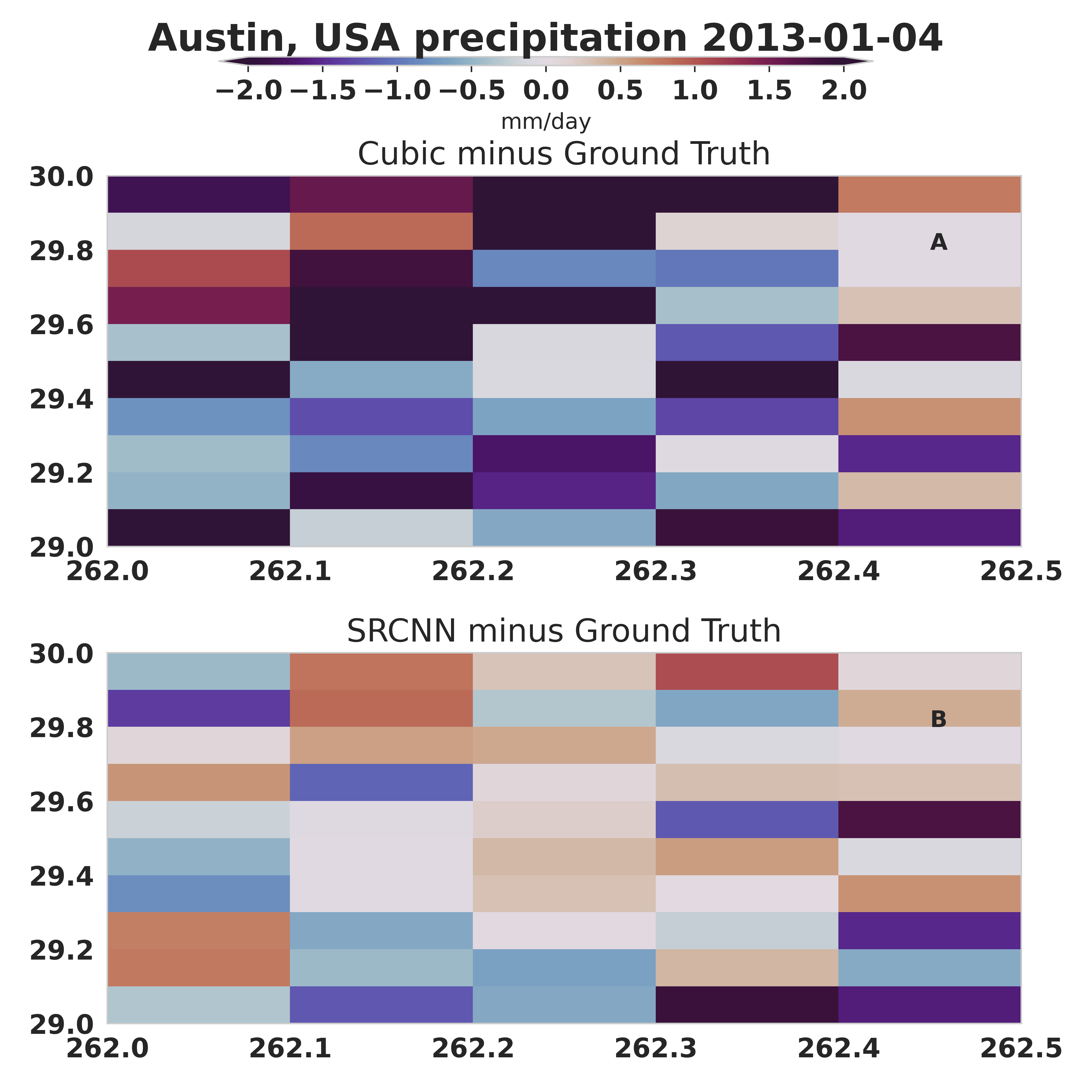}
{\caption{Difference in precipitation maps as cubic interpolation minus the JAXA GsMAP satellite rainfall, and SRCNN minus the JAXA GsMAP satellite rainfall}
\label{fig1}}
\end{center}
\end{figure}

\subsection{Study contributions}

In this study, we employ a deep CNN-based model named SRCNN to iteratively downscale the precipitation data over Austin, USA, from JAXA global product available at 0.1 hourly spatiotemporal resolutions from 2000 to 2020. We take the inspiration from the general purpose statistical downscaling operators such as cubic or linear interpolation, which can theoretically increase the resolution to generate finest-scale datasets. The task performed by the traditional statistical operators is to rearrange the low resolution information into a dense matrix. We attempt similar transformation and develop the iterative SRCNN to perform 2x downscaling at any scale, i.e., from 10km to 5km, from 5km to 2.5 km and so on. Thus, our contributions include the development of a general-purpose algorithm based on a single image super-resolution in computer vision by using sub-images to perform iterative downscaling for high-resolution urban-scale datasets. Our approach can be used to iteratively generate downscaled products, theoretically up to any spatial resolution for which images are available. As stated, we focus on achieving spatial resolutions of $\sim$ 300 meters. This spatial resolution can be used for targeted applications to communities planning climate resiliency and adaptation strategies for urban neighborhoods. Generating $\sim$ 300-meter precipitation maps from measurements is challenging as it would require the deployment of measurement sensors in large quantities. The initial setup of such sensors and their regular maintenance also would be quite expensive for initial setup and even more challenging for maintaining and operating. The iterative method provides a workaround for the traditional techniques by generating high-resolution urban precipitation datasets at a low cost. We target a scaling factor of around 30x from input low-resolution satellite product at 0.1 degrees or ~10km to a target high-resolution output of $\sim$ 300 meters.

\section{Data and methodology}

This section discusses the data and methodology used in this study. The schematic of the data and methodology in addition to the applications of high resolution urban datasets is shown in Figure 2.

\subsection{Dataset}
The precipitation dataset used in this study is derived from the Global Rainfall Watextbf (GSMaP) algorithm of Japan Aerospace Exploration Agency(JAXA). JAXA GsMAP is a part of the Global Precipitation Measurement (GPM) which started on 28 February 2014 as a successor to Tropical Rainfall Measuring Mission (TRMM). The state-of-the-art Ku/Ka Doppler dual-frequency precipitation radar (DPR) and microwave imager are aboard GPM's primary satellite. Because of the improvements made by its load over TRMM in identifying tropical precipitation (0-1 millimetres per day) (\cite{draper2015global}), GPM provides hydrological researchers more precise global satellite precipitation measurements than were possible with TRMM. GPM consists of two algorithms for satellite based precipitation. One of them is the IMERG (\cite{hou2014global}) from NASA, and the other is GsMAP (\cite{kubota2007global}) from JAXA. JAXA GsMAP has various products in its catalogue, viz, near real time, moving vector with Kalman filter and the gauge calibrated standard product. We have used the JAXA GsMap gauge-calibrated product for this study. The dataset provides global coverage and is available from 2000 to present at a spatial resolution of 0.1$^{\circ}$ and a temporal resolution of one hour. A rectangular box consisting of the subdomain of JAXA data over Austin, USA, covering 29-32N, 96-99W was selected. 

Although a large-scale dataset over the Earth needs to consider sphericity, our dataset is over a small region relative to the global dataset from which it is acquired. Because of the smaller domain size, it is considered as a two-dimensional image. The algorithms applicable to two-dimensional images in computer vision are therefore considered suitable for this data. The data is first split into training and testing data with the train data corresponding to 2001 to 2009 and test data as 2010 to 2018. The domain over Austin is selected as a 3 $^{\circ}$ x 3 $^{\circ}$ box and the spatial resolution of JAXA GsMAP data is 0.1 $^{\circ}$ x 0.1 $^{\circ}$. Thus, the original data is a matrix of size 30 x 30. 
\subsection{Methodology}
The training data is normalized using min-max scaling and then transformed to sub-images of size 20 × 20. The test dataset is scaled using the normalization weights from the training data. The sub-images are then fed into the SRCNN algorithm. The deep convolutional neural network employs the following equations: 
\begin{equation}
F1(Y) = max(0, W1 * Y + B1)
\label{eq1}
\end{equation}
\begin{equation}
F2(Y) = max(0, W2 * F1(Y) + B2)
\label{eq2}
\end{equation}
\begin{equation}
F3(Y) = max(0, W3 * F2(Y) + B3)
\label{eq3}
\end{equation}

The architectural details of the model can be seen in detail from \cite{dong2015image}. First, the 20 × 20 sub-images are convoluted by 64 filters, each with a size 9 × 9 and then the rectified linear unit (RELU) activation function is operated upon the convolutions. This operation involves the convolution and non-linear activation functions described by Equation 1. The output of equation 1 again goes through the transformation involving convolution and activation, however, 32 filters of size 1 × 1 complete the operations of Equation 2. The output of equation 2 is then operated by one 5 × 5 filter to perform a linear combination operation which is the output of equation 3. We use the padding option ‘same’ so that the operations are padded, and the size of input and output remains identical. The applied optimizer is adaptive moment estimation (ADAM) with a learning rate of 0.001. The model is trained for the years 2001 to 2008, with the year 2009 corresponding to the validation period during training. The best model is saved every 100 iterations if the validation loss falls below the previous best model. Once the training is completed, the test data is first normalized and broken down into sub-images to be fed as an input to the trained model. The test predictions are then reconstructed from the sub-images and inverse normalized to compare with the ground truth in the test dataset. The model is a general purpose operator which is capable of performing twofold super-resolution from any low resolution information to a higher resolution matrix. 

In a nutshell, the output from SRCNN at ~ 10 km spatial resolution is used as an input to the trained model to generate outputs at ~ 5 km in an iterative manner. This is possible as we train the model using sub-images which is the core of our algorithm. Hence the model is agnostic to the size of the input dataset and any large image can be broken into sub-images of size 20 x 20. These sub-images are reconstructed back after the model predictions. This property of our deep learning-based solution makes the model comparable to a standard interpolation technique such as cubic interpolation. Thus, it is henceforth used in an interative framework to generate the urban scale ($\sim$ 300 m) product from 10 km JAXA GsMAP satellite precipitation product. The complete training with early stopping took around one week on an NVIDIA Tesla P100 GPU. The super-resolution downscaled datasets are generated in an hour from the trained model for the entire period of ten years of testing data.
\begin{figure}[t]%
\begin{center}

\includegraphics[width=0.55\textwidth]{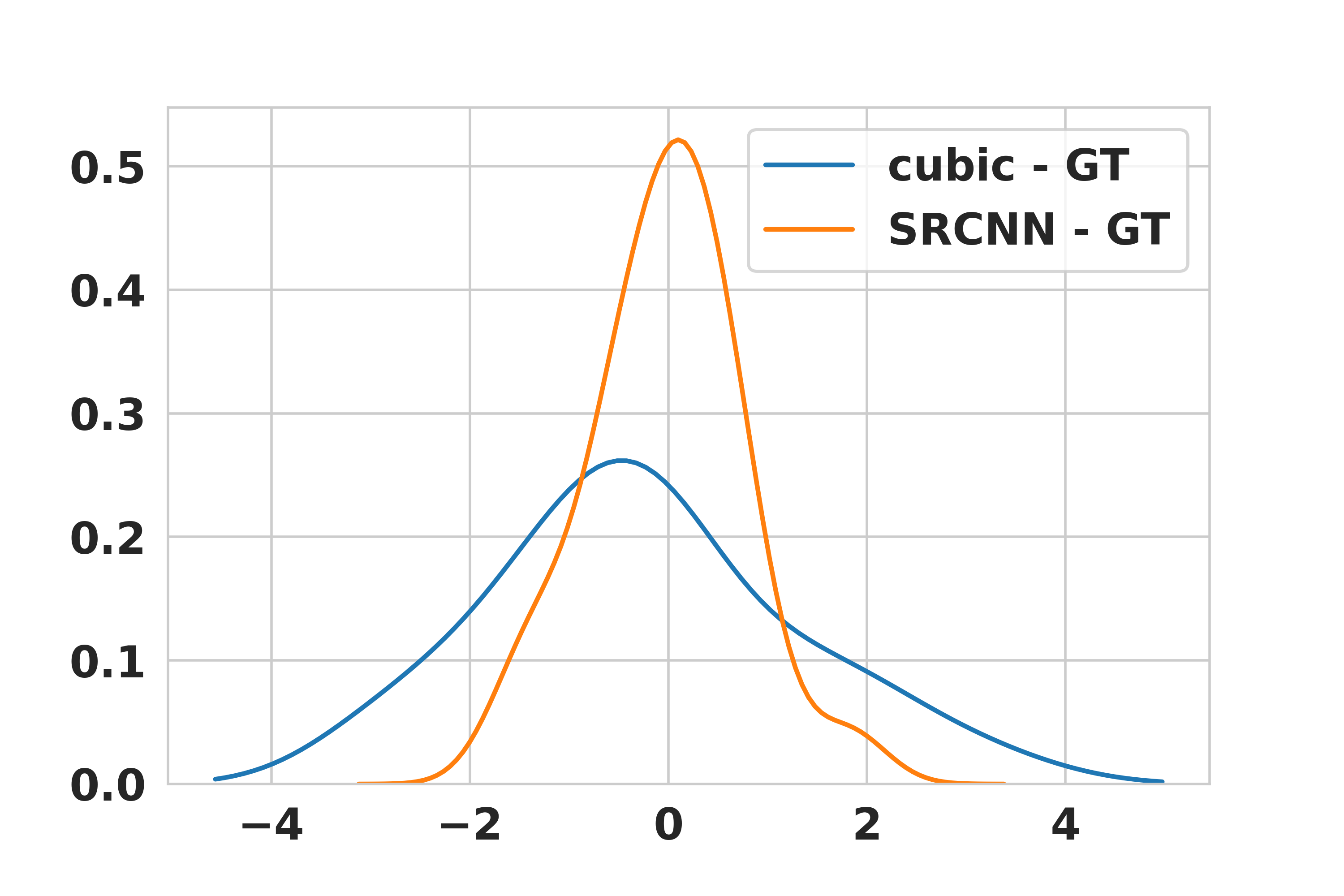}
{\caption{Distributions of cubic minus ground truth (GT) and SRCNN minus ground truth (GT) show under and overestimation of precipitation in cubic interpolation baseline}
\label{fig1}}
\end{center}
\end{figure}

\begin{figure}[t]%
\begin{center}

\includegraphics[width=0.55\textwidth]{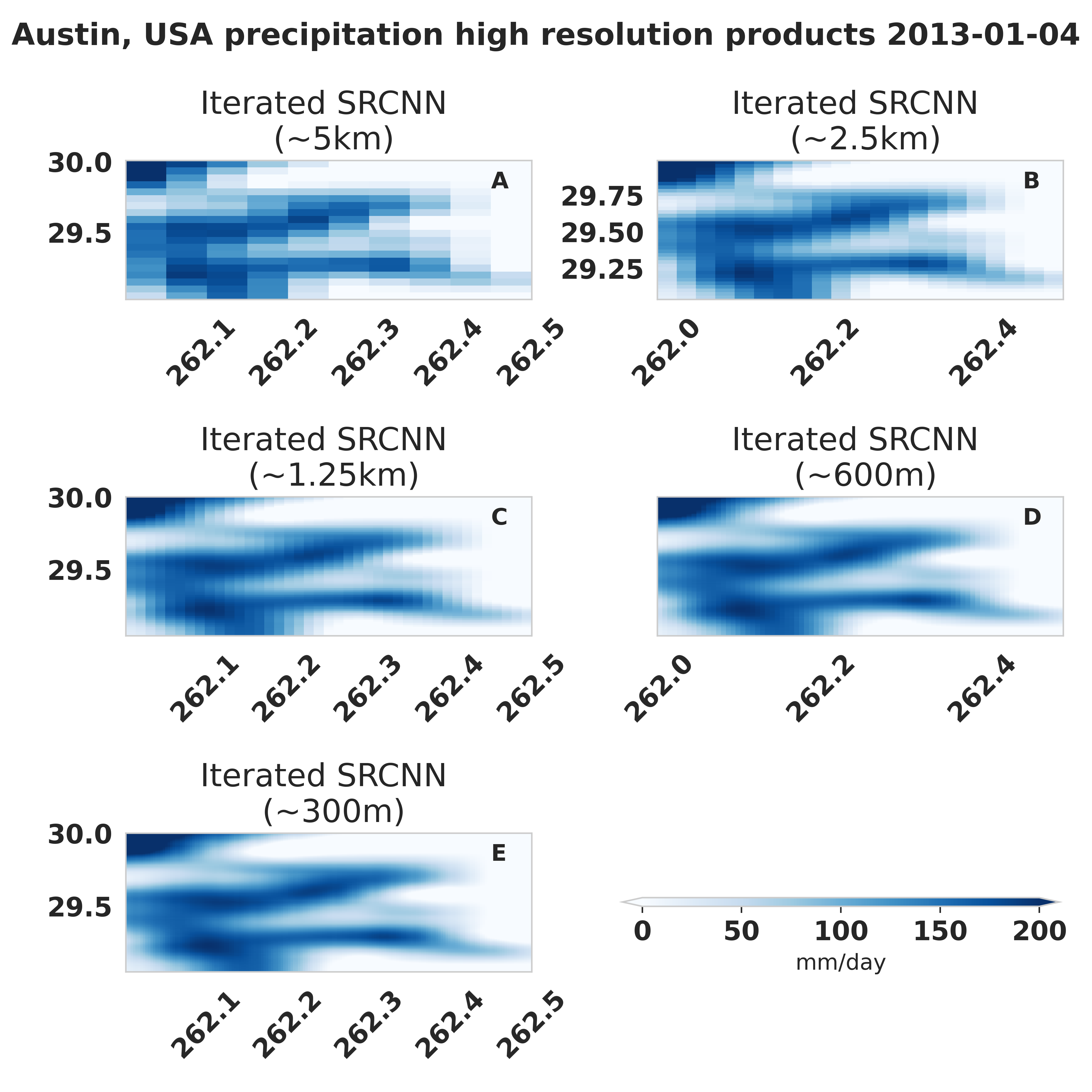}
{\caption{High-resolution maps of the gridded rainfall product over Austin using Iterated SRCNN. The resolution upto 300 m is generated by the method.}
\label{fig1}}
\end{center}
\end{figure}

\section{Results and discussion}
The test predictions generated for the period 2010 to 2019 are compared with the ground truth data for the same period. The hourly data that report no rainfall in the input is masked out from the metrics used to compare the cubic interpolation baseline and SRCNN based deep learning model. The statistical results (shown in Table 2) denote the improvements acquired using the SRCNN method. Two metrics, viz, Peak Signal to Noise Ratio (PSNR) and Mutual Information are used to compare the entire test predictions with the ground truth, which is the JAXA GsMAP satellite product. PSNR quantifies the ratio of a signal's highest potential strength to its corrupting noise power. It is measured by the decibel scale. PSNR is a typical metric for gauging the quality of lossy image. Compression introduces pixel level errors into the data and PSNR is an approximation of human perception of reconstruction quality when comparing compression codecs (\cite{li2007robust}). Thus PSNR is a suitable metric to compare pixel or grid level errors in two images or matrices. Of the other two evaluation metrics, and mutual information is an advanced nonlinear index measuring the similarity of distributions. It has also been called the correlation of the 21st century (\cite{speed2011correlation}). The PSNR increases from 146.96 in the baseline to 149.46 for SRCNN compared with the original 10 km dataset. In addition, the rainfall maps generated with the SRCNN approach increased mutual information from 0.59 (in the baseline) to 0.62. 

\begin{table}[t]
\begin{center}

\tabcolsep=3pt
\caption{Performance of cubic interpolation and iterative SRCNN used in this study}\label{tab2}%
\begin{tabular}{@{}lcc@{}}
\toprule
\textbf{METRIC} & \textbf{CUBIC INTERPOLATION (BASELINE)} & \textbf{SRCNN}    \\
\midrule
\textbf{PSNR} & 146.96 & 149.46 \\
\textbf{MUTUAL INFORMATION} & 0.59 & 0.62 \\

\hline

\end{tabular}

\end{center}
\end{table}

A spatial comparison of SRCNN prediction and the cubic interpolation baseline with the ground truth is shown in Figure 3 as the 4th of January 2013 daily precipitation. These matrices show improved pixel-level information in SRCNN. Although, our algorithm shows pixel level improvements of up to 2 mm/day at the resolution of the input satellite data, the high resolution precipitation product is a matrix of size 30x relative to the input satellite data (10 km). Small errors averaged over a large 10 km X 10 km grid would exponentially grow to a large bias in a concentrated gridded product of grid size 300 m X 300 m. Iterative forecasts are generated up to ~300 m and beyond ~5 km iterative forecasts. Visual improvements in the downscaled urban data can be seen from Figure 3. We show a sample prediction for a heavy precipitation event that impacted over Austin on the 4th of January 2013 at 8 pm.

Figure 4. shows under and over representation of the pixel level rainfall amount can be observed in the cubic interpolation relative to SRCNN. In addition, horizontal and vertical cutting lines are visible in the downscaled outputs from deep learning. These cutting lines are an artefact of SRCNN and have been previously identified (\cite{song2019improved}). The use of advanced deep learning-based super-resolution methodologies for downscaling will free the outputs from the cutting lines. Figure 5. shows the spatial maps of the product at 300 m X 300 m gridded spatial resolution. It can be visually seen that the rainfall becomes smooth as the resolution increases. Zooming over specific regions would entail the differences at different resolutions.

\section{Conclusions and future work}

In this study, a deep learning-based convolutional neural network also known as SRCNN is employed to generate high-resolution maps of precipitation at 300 meters from the satellite-based JAXA GsMAP product over Austin, USA (29-32N, 96-99W). The deep learning-based model is trained for the period 2001 to 2009 and test predictions are  generated for the period 2010 to 2019. The quantitative metrics, as well as visual inspection, show substantial improvements in the pixel level information for the SRCNN relative to the baseline cubic interpolation. The highest resolution dataset at 300 meters is generated by using iterative prediction. 

The dataset can be further improved by improving the trained model using hyperparameter tuning as a follow-up study. Increasing the number of layers and using residual or generative networks can further improve. It can also be built into a solution that takes in the coarse-resolution satellite data as an input and returns the high-resolution output in real-time. This software-as-a-service paradigm can be used to aid the flood, typhoon, hurricane or cyclone-affected areas in real-time, similar to the real-time JAXA global rainfall watextbf (https://sharaku.eorc.jaxa.jp/GSMaP/index.htm).

Our framework provides a efficient model with low computational cost and higher speed (compared to other CNN methods) for generating high-resolution data over urban regions that could be used by city planners for smarter and more efficient developments.  To the best of our knowledge, this is the first study to perform urban downscaling for precipitation datasets. The increasing trend of extreme rainfall and flash floods over urban areas make our high-resolution spatiotemporal dataset critical to identify potentially vulnerable areas and address the infrastructural requirements of those regions. Advanced  deep learning algorithms using attention and generative learning would be required for further improving these high-resolution products.

\section*{Acknowledgments}
The authors thank Harsh Kamath and Ting-Yu Dai for their constructive comments.

\bibliographystyle{unsrt}  
\bibliography{references}

\end{document}